\makeatletter
\declare@file@substitution{revtex4-1.cls}{revtex4-2.cls}
\makeatother

\documentclass[preprint]{aastex631}

\usepackage{amsmath}
\usepackage{dsfont}
\usepackage{bm}
\usepackage{hyperref}
\newcommand{\al}[1]{{#1}}

\begin{document}

\title{The importance of Berry phase in solar acoustic modes}

\author[0000-0001-6136-4164]{Armand Leclerc}
\altaffiliation[Corresponding author: ]{armand.leclerc@ens-lyon.fr}
\affiliation{ENS de Lyon, CRAL UMR5574, Universite Claude Bernard Lyon 1, CNRS, Lyon, F-69007, France}
\author{Guillaume Laibe}
\affiliation{ENS de Lyon, CRAL UMR5574, Universite Claude Bernard Lyon 1, CNRS, Lyon, F-69007, France}
\affiliation{Institut Universitaire de France}

\begin{abstract}
An analytic expression for the frequencies of standing waves in stars, applicable to any radial order $n$, 
is derived from ray-tracing equations by the mean of Wigner-Weyl calculus. A correction to previous formulas 
currently employed in asteroseismology is identified as the Berry phase, which accounts for the vectorial nature 
of wave propagation in stars. Accounting for this quantity significantly improves upon previous laws for low $n$ 
modes of the Sun, and we show that the Berry phase is indeed present in the available observational data of 
solar modes. This phase is due to inhomogeneities of the medium.
\end{abstract}


\section{Introduction}

Mechanical waves that travel through the interiors of stars provide the most accurate insights into their inner structures (e.g. \citealt{Basu2016,JCD2021}). Since these waves manifest as standing oscillations on the surface, a standard approach consists in deriving the equations that describe standing waves to obtain the theoretical frequency distributions of pulsations for large radial orders ($n \gg 1$). The frequencies are expressed as functions of the harmonic degree $\ell$ and the physical parameters of the star: the sound speed $c_{\rm s}$ and the buoyant frequency $N$, which both vary with radius $r$. 
Acoustic modes follow a relationship known as Duvall's law: frequencies are distributed approximately uniformly, as given by the formula ${\nu = (n + \frac{\ell}{2} + \frac{3}{4})\Delta\nu}$, where ${\Delta\nu = (2\int_0^R \mathrm{d}r/c_\mathrm{s})^{-1}}$ is the so-called large frequency separation \citep{duvall1982}. Gravity modes obey Tassoul's law: periods are approximately uniformly spaced and expressed as ${P = \frac{2\pi^2(n+\frac{1}{2})}{\sqrt{\ell(\ell+1)}} ( \int_{r_1}^{r_2}\frac{N}{r}\,\mathrm{d}r )^{-1}}$ \citep{shibahashi1979,tassoul1980}. 
These laws give explicit constrain on $N$ and $c_\mathrm{s}$ from the observed oscillation frequencies. At large radial orders $n$, deviations from these distributions are caused by the acoustic cut-off frequency ${\omega_\mathrm{c}^2 = c_\mathrm{s}^2/4H^2(1-2\mathrm{d}H/\mathrm{d}r)}$, where $H$ is the pressure scale height. For acoustic modes, these corrections are usually grouped into a parameter called the phase function $\alpha(\omega)$ (e.g.  \citealt{deubner1984,christensen1992}), which leads to a small frequency separation ${\delta\nu = \nu_{n,\ell}-\nu_{n+1,\ell-2}}$, offering valuable information about the temperature gradient of the star \citep{tassoul1980}. 

In this study, we provide a more precise equation for the frequencies of standing waves that is valid for all radial orders. Importantly, this equation introduces a correction term to the previously established laws that improves significantly their accuracy for low $n$. It is possible to relax the usual high $n$ approximation by relying instead on a high $\ell$ approximation. For this, we treat asteroseismology from the alternate perspective of geometrical optics, which describes the paths of wave rays \citep{gough1986,gough1993,loi2020}. While ray-tracing is generally based on a scalar description of the wave, we account here for polarization effects due to the multicomponent nature of the perturbations in the equations, resulting in a more accurate formulation. This result is obtained by treating the full perturbations equations as a Schrödinger equation, and taking its semi-classical limit which yields the dynamics of $p$-quasi-particles and $g$-quasi-particles.
\begin{figure}
    \centering
    \includegraphics[width=0.45\columnwidth]{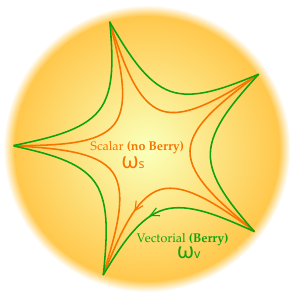}
    \includegraphics[width=0.45\columnwidth]{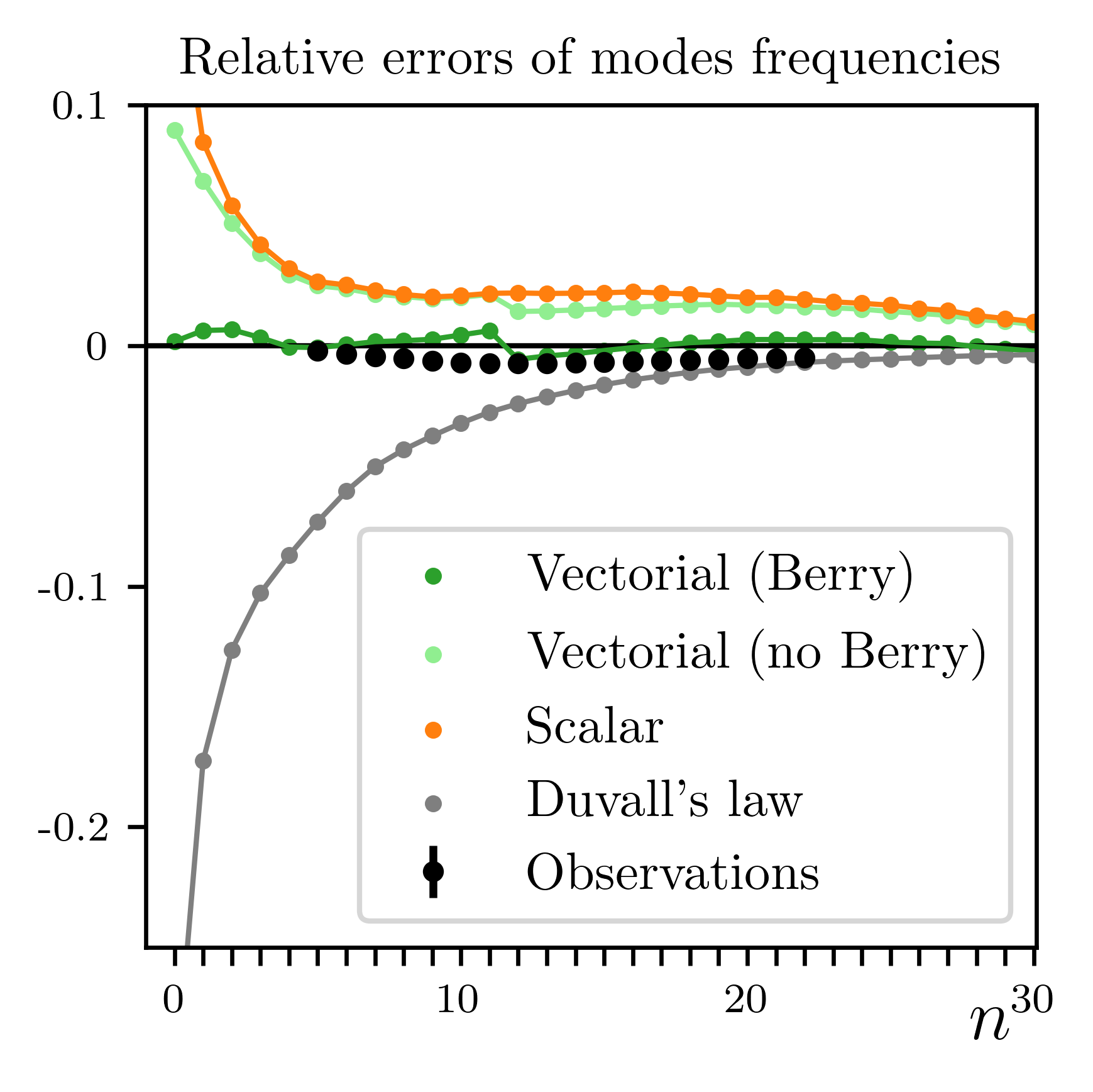}
    \caption{Ray-tracing equations accounting for the Berry curvature are slightly deviated, correcting the frequencies of standing waves. \textbf{Left:} The polarization of the wave adds a Berry term in the ray-tracing equations, changing the bending of the ray between a scalar and fully vectorial theory. The resulting standing wave traced by the rays have a different frequency. \textbf{Right:} Frequency estimates of solar acoustic modes for azimuthal degree  $\ell = 25$ significantly improve when accounting for the Berry phase, in particular for low radial orders $n$. The scalar and the two  vectorial theories account for the phase function $\alpha(\omega)$, the "vectorial (no Berry)" theory accounts for correction of $\Omega$ arising from spatial variation of background quantities, the "vectorial (Berry)" theory further incorporate $\phi_\mathrm{B}$. For $n\leq20$, only the predictions accounting for $\phi_\mathrm{B}$ match the observed frequencies to the current degree of precision of the 1D model (1\%). \al{The observational uncertainties (vertical bars) are undistinguishable at this scale ($0.1\%$ in \citep{hill1996}).}}
    \label{fig:soleil}
\end{figure}

 \section{Ray-tracing with varying polarizations}

The evolution of linear, adiabatic perturbations of a non-rotating, non-magnetic stars under the Cowling approximation can be expressed in a symmetric form by adopting the appropriate change of variables
\begin{eqnarray}
    \bm{v} &=& \rho_0^{1/2}c_\mathrm{s}^{1/2} r \bm{v}^\prime,\\
    p &=& \rho_0^{-1/2}c_\mathrm{s}^{-1/2}r p^\prime,\\
    \Theta &=& \rho_0^{-1/2}r \frac{g}{N}(\rho^\prime - \frac{1}{c_\mathrm{s}^2}p^\prime),
\end{eqnarray}
where $p'$, $\rho'$ and $\bm{v}'$ are the Eulerian perturbations in pressure, density and velocity. It slightly differs here from the changes of variables performed in \citep{leclerc2022,leclerc2024}, in which the detailed derivations are presented.\\
Further decomposing the velocity perturbation onto the basis of vectorial spherical harmonics, one has 
\begin{equation}
    \bm{v}(r,\theta,\phi) = v_r(r)\bm{Y}_\ell^m + v_h(r)\bm{\Psi}_\ell^m,  
\end{equation} 
where $\bm{Y}_\ell^m = Y_\ell^m \mathbf{e}_r$ and $\bm{\Psi}_\ell^m = \frac{ir}{\sqrt{\ell(\ell+1)}}\bm{\nabla}Y_\ell^m$. 
By introducing the acoustic radius ${z_\mathrm{s}}$ defined by
$\mathrm{d}{z_\mathrm{s}} = \mathrm{d}r/c_\mathrm{s}$ as a new radial coordinate, the equations of perturbation read
\begin{eqnarray}
    &&i\partial_t \begin{pmatrix}
        v_h \\ v_r \\ \Theta \\ p
    \end{pmatrix} = \begin{pmatrix}
        0 & 0 & 0 & L_\ell \\
        0 & 0 & -iN & -i\partial_{z_\mathrm{s}} + iS \\
        0 & iN & 0 & 0 \\
        L_\ell & -i\partial_{z_\mathrm{s}} - iS & 0 & 0
    \end{pmatrix}\begin{pmatrix}
        v_h \\ v_r \\ \Theta \\ p
    \end{pmatrix}.
\end{eqnarray}\\
Defining the slow acoustic radius and slow time as $z = \epsilon z_\mathrm{s}$, $\tau = \epsilon t$ with $\epsilon = 1/\sqrt{\ell(\ell+1)}$, one obtains the wave equation
\begin{eqnarray}
    && i\epsilon \partial_\tau \bm{X} = \hat{\mathbf{H}} \bm{X},
    \label{eq:schrodinger_equation}\\
    &&\text{with}\quad\hat{\mathbf{H}} = \begin{pmatrix}
        0 & 0 & 0 & L_\ell \\
        0 & 0 & -iN & -i\epsilon\partial_z + iS \\
        0 & iN & 0 & 0 \\
        L_\ell & -i\epsilon\partial_z - iS & 0 & 0
    \end{pmatrix} ,
    \label{eq:op_H}
\end{eqnarray}
where $\bm{X} = \begin{pmatrix}v_h & v_r & \Theta & p\end{pmatrix}^\top$, $L_\ell = c_\mathrm{s}/\epsilon r$ and $N^2 = g(\frac{1}{\Gamma_1}\frac{\mathrm{d}\ln P_0}{\mathrm{d}r} - \frac{\mathrm{d}\ln\rho_0}{\mathrm{d}r})$ are the Lamb and the buoyancy frequencies and $S = \frac{c_\mathrm{s}}{2g}\left( N^2 - \frac{g^2}{c_\mathrm{s}^2} \right) - \frac{1}{2}\frac{\mathrm{d}c_\mathrm{s}}{\mathrm{d}r} + \frac{c_\mathrm{s}}{r}$. The parameter $\epsilon \equiv 1/\sqrt{\ell(\ell+1)}$ is the angular wavelength which acts as a small parameter for large azimuthal degrees $\ell$. \\
In principle, other small parameters $\epsilon$ can be chosen, as long as the limit $\epsilon\to0$ is a limit in which the frequencies of $p$-modes and $g$-modes are well-separated which is needed in the following. \al{This choice of variables $\begin{pmatrix}v_h & v_r & \Theta & p \end{pmatrix}^\top$ and coordinate $z$ yields an operator which is self-adjoint with respect to the scalar product $\langle \bm{X}_1,\bm{X}_2 \rangle = \int \mathrm{d}z\;\bm{X}_1^\dagger\cdot\bm{X}_2$. The expressions of our results are obtained for those specific variables and coordinate.}

\al{Obtaining the equations of rays from Eq.~\eqref{eq:op_H} at order $\epsilon^1$ is involved and requires a number of technical steps.} The derivation is analogous to the ones of \citealt{perez2021} and \citealt{venaille2023}. We first aim to transform the vectorial equation~\eqref{eq:op_H} to a scalar one by transforming it to the form
\begin{equation}
    i\epsilon\partial_\tau \psi = \hat{{\Omega}}\psi ,
\end{equation}
and this, for a selected waveband (either acoustic or internal gravity). The idea is to reconstruct the multicomponent perturbations field $\bm{X}(\tau,z)$ is  from the scalar field $\psi(\tau,z)$ through a vectorial operator $\hat{\bm{\chi}}(z,\partial_z)$ by $\bm{X} = \hat{\bm{\chi}}\psi$. $\psi$ is the scalar field that evolves according to the dispersion relation, and $\hat{\bm\chi}$ is a vector of differential operators used to reconstruct all perturbed fields of the wave from the scalar field. In general, such transform is not possible. It become however feasible when $\epsilon \ll 1$, which ensures that the wave bands are well-separated. This section shows that the ray-tracing equations for $\bm{X}$ and $\psi$ differ slightly, by a term involving the polarization relations given by $\hat{\bm{\chi}}$.\\
The expressions of $\hat{\Omega}$ and $\hat{\bm{\chi}}$ are unknown and have to be determined. The condition $\hat{\bm{\chi}}^\dagger\cdot\hat{\bm{\chi}} = \mathds{1}$ is imposed, so that the energy of the wave is $E = \int\mathrm{d}z\; \bm{X}^\dagger \cdot \bm{X} = \int\mathrm{d}z\; \psi^*\psi = 1$. The symbol ${}^\dagger$ denotes the conjugate-transpose in the sense of $4\times4$ complex matrices and vectors. Consistency of time evolution imposes
\begin{equation}
    \hat{\mathbf{H}}\hat{\bm{\chi}} = \hat{\bm{\chi}}\hat{\Omega}.\label{eq:def-chi-Omega}
\end{equation}
The operators are then expanded to first order in $\epsilon$ as
\begin{eqnarray}
    \hat{\mathbf{H}} &=& \hat{\mathbf{H}}_0 + \epsilon\hat{\mathbf{H}}_1,\\
    \hat{\Omega} &=& \hat{\Omega}_0 + \epsilon\hat{\Omega}_1,\\
    \hat{\bm{\chi}} &=& \hat{\bm{\chi}}_0 + \epsilon\hat{\bm{\chi}}_1.
\end{eqnarray}
We now apply the Wigner transform, which  transforms a differential operator $\hat{A}(z,\partial_z)$ into a function on the phase space $A(z,k_z)$. It is the reciprocal transform of the Weyl transformation, which is a quantization rule. See \citep{onuki2020} for a detailed description of the Wigner transform and its usefulness in fluid mechanics. For instance, the Wigner transform gives the maps 
\begin{eqnarray}
    -i\partial_z &\mapsto& k_z,\\
    c_\mathrm{s}(\hat{z}) &\mapsto& c_\mathrm{s}(z).
\end{eqnarray}
It is a way of representing the local action of operators on plane waves. Wigner-Weyl calculus comes with a general way of computing products of operators, the so-called Moyal product $\star$, such that
\begin{equation}
    \hat{A}\hat{B} = \widehat{{A\star B}} ,
\end{equation}
which is particularly useful to treat Eq.~\eqref{eq:def-chi-Omega}. The Moyal product is defined as the expansion on $\epsilon$ \citep{onuki2020}
\begin{equation}
    A\star B \equiv \sum_{p,q\in\mathds{N}^2}\frac{(-1)^p}{p!q!}\left(\frac{i}{2}\epsilon\right)^{p+q}(\partial_z^q \partial_{k_z}^p A)(\partial_z^p \partial_{k_z}^q B),
\end{equation}
which gives at first order in $\epsilon$ the r.h.s of Eq.~\eqref{eq:def-chi-Omega} as
\begin{eqnarray}
    \hat{\bm{\chi}}\hat{\Omega} &=& \widehat{\bm{\chi}\star \Omega}\\
    \mathrm{and}\;\bm{\chi}\star \Omega &=& \bm{\chi}_0\Omega_0 + \epsilon\bigg(\bm{\chi}_1\Omega_0 + \bm{\chi}_0\Omega_1 + \frac{i}{2}\{\bm{\chi}_0,\Omega_0\}\bigg), \label{eq:moyal_exp}
\end{eqnarray}
where $\{,\}$ are the Poisson bracket in the phase space $(z,k_z)$ defined in the main text.\\
From Eq.~\eqref{eq:moyal_exp}, one determines the symbols $\Omega_0$, $\Omega_1$ and $\bm{\chi}_0$. At zeroth order in $\epsilon$, one has
\begin{equation}
    \mathbf{H}_0\Omega_0 = \bm{\chi}_0\Omega_0 ,
\end{equation}
where $\Omega_0$ and $\bm{\chi}_0$ are an eigenvalue and an eigenvector of the matrix $\mathbf{H}_0$, as obtained with direct linear algebra. The matrix $\mathbf{H}_0$ reads
\begin{equation}
    \mathbf{H}_0(z,k_z) = \begin{pmatrix}
        0 & 0 & 0 & L_\ell \\
        0 & 0 & -iN & k_z + iS \\
        0 & iN & 0 & 0 \\
        L_\ell & k_z - iS & 0 & 0
    \end{pmatrix}.
\end{equation}
The full expressions of $\chi_0$, $\Omega_0$ are given in Appendix~\ref{app:berry-curv} for both acoustic and internal gravity waves.\\
Exploiting the limit $\epsilon \ll 1$, Wigner-Weyl calculus has therefore provided the mathematical framework to project the operator $\hat{\mathbf{H}}$  onto the acoustic band, by separating the scalar propagation operator $\hat{\Omega}$ and the polarisation operator $\hat{\bm{\chi}}$. Pursuing the expansion of Eq.~\eqref{eq:def-chi-Omega} at order $\epsilon^1$, one finds
\begin{equation}
    \Omega_1 = \frac{i}{2} \bm{\chi}_0^\dagger\{\mathbf{H}_0 - \Omega_0 \mathbf{I}_4,\bm{\chi}_0\} + \frac{i}{2} \bm{\chi}_0^\dagger\{\Omega_0 \mathbf{I}_4,\bm{\chi}_0\}.\label{eq:omega_1}
\end{equation}
The first term of the r.h.s is involved in the ray-tracing dynamics. The second term is not involved in the ray-tracing equations \citep{perez2021,venaille2023}.\\

We can now establish the time evolution of a wavepacket of the form $\bm{X}(\tau,z) = \hat{\bm{\chi}} (z,k(\tau,z)) \psi (\tau,z)$, where
\begin{equation}
    \psi(\tau,z) = a_0(\tau,z)\mathrm{e^{\frac{i}{\epsilon}(\phi_0+\epsilon\phi_1)}}.
\end{equation}
The envelope $a_0$ is chosen to have has significant values within a narrow spatial region $\Delta z$ that is small compared to any other length-scale in the star. From \citet{venaille2023}, one has at order $\epsilon$ 
\begin{eqnarray}
    \bm{X} &=& a_0\mathrm{e^{\frac{i}{\epsilon}(\phi_0+\epsilon\phi_1)}}\bm{\chi}_0(z,k(\tau,z)),\\
    k(\tau,z) &\equiv& \partial_z\phi_0 + \epsilon\partial_z\phi_1.
\end{eqnarray}
The average position and momentum of the wavepacket are then determined by $a_0$ and $\phi_0+\epsilon\phi_1$ respectively. The coordinates of the scalar wavepacket in phase space are
\begin{eqnarray}
    \langle z\rangle_\psi &\equiv& \int\mathrm{d}z\;\psi^* z\psi = \int\mathrm{d}z\;z\, a_0^2,\\
    \langle k_z\rangle_\psi &\equiv&\int\mathrm{d}z\;\psi^* (-i\epsilon\partial_z)\psi = (\partial_z\phi_0 + \epsilon\partial_z\phi_1)\big\vert_{z=\langle z\rangle_\psi}.
\end{eqnarray}
Taking the derivative with respect to time, using the property that $\hat{\Omega}$ is self-adjoint and the identities $\hat{z}\hat{\Omega} - \hat{\Omega}\hat{z} = i\epsilon \widehat{\partial_{k_z}\Omega}$ and $\partial_z\hat{\Omega} - \hat{\Omega}\partial_z = \widehat{\partial_{z}\Omega}$ \citep{venaille2023}, one obtains the dynamical evolution
\begin{eqnarray}
    \dot{\langle z\rangle}_\psi &=& \int \mathrm{d}z\,(\partial_\tau \psi^*\cdot \hat{z} \psi + \psi^*\cdot \hat{z} \partial_\tau \psi),\\
     &=& \int \mathrm{d}z\,\frac{i}{\epsilon}( \psi^*\cdot \hat{\Omega}\hat{z} \psi - \psi^*\cdot \hat{z}\hat{\Omega} \psi),\\
     &=& \int \mathrm{d}z\,\psi^* \widehat{\partial_{k_z}{\Omega}} \psi,
\end{eqnarray}
and 
\begin{eqnarray}
    \dot{\langle k_z\rangle}_\psi &=& \int \mathrm{d}z\,(\partial_\tau \psi^*\cdot (-i\epsilon\partial_z) \psi + \psi^*\cdot (-i\epsilon\partial_z) \partial_\tau \psi),\\
     &=& \int \mathrm{d}z\,( \psi^*\cdot \hat{\Omega}\partial_z \psi - \psi^*\cdot \partial_z\hat{\Omega} \psi),\\
     &=& -\int \mathrm{d}z\,\psi^* \widehat{\partial_{z}{\Omega}} \psi.
\end{eqnarray}
Using the general result that for a scalar wavepacket of small extension and any operator $\hat{A}$, one has $\int\mathrm{d}z \;\psi^*\hat{A}\psi \sim A(\langle z\rangle_\psi,\langle k_z\rangle_\psi)$ \citep{venaille2023}, one obtains for the scalar wavepacket
\begin{eqnarray}
    \dot{\langle z\rangle}_\psi &=& +\partial_{k_z}\Omega,\label{eq:zdot_scalar_wavepacket}\\
    \dot{\langle k_z\rangle}_\psi &=& -\partial_{z}\Omega.\label{eq:kzdot_scalar_wavepacket}
\end{eqnarray}
This is a canonical system for the Hamiltonian $\Omega = \Omega_0+\epsilon\Omega_1$. \citet{littlejohn1991} showed that this expression is not gauge-invariant under a change of global phase $\bm{\chi}_0 \mapsto \mathrm{e}^{i g(z,k_z)}\bm{\chi}_0$ due to the last term in $\Omega_1$ in Eq.~\eqref{eq:omega_1}. On the other hand, the coordinates of the vectorial wavepacket
\begin{eqnarray}
    \langle z\rangle_{\bm{X}} &\equiv& \int \mathrm{d}z\,\bm{X}^\dagger\cdot z \bm{X},\\
    \langle k_z\rangle_{\bm{X}} &\equiv& \int \mathrm{d}z\,\bm{X}^\dagger\cdot  (-i\epsilon\partial_z)\bm{X},
\end{eqnarray}
are necessarily gauge-independent of the choice of phase of $\bm{\chi}_0$, as they are defined independently of the decomposition $\bm{X}= \hat{\bm{\chi}}\psi$. The evolution of these coordinates can be obtained from the ones of the scalar wavepacket by the mean of the relations
\begin{eqnarray}
    \langle z\rangle_{\bm{X}} &=&\langle z\rangle_\psi + i\epsilon \bm{\chi}_0^\dagger\cdot\partial_{k_z}\bm{\chi}_0,\label{eq:change-coordinate-1}\\
    \langle k_z\rangle_{\bm{X}} &=& \langle k_z\rangle_\psi - i\epsilon\bm{\chi}_0^\dagger\cdot\partial_{z}\bm{\chi}_0 .\label{eq:change-coordinate-2}
\end{eqnarray}
One observes that the last terms of the right-hand side of Eqs.~\eqref{eq:change-coordinate-1}-\eqref{eq:change-coordinate-2} are not gauge-independent as they would change under $\bm{\chi}_0 \mapsto \mathrm{e}^{i g(z,k_z)}\bm{\chi}_0 $, implying that the coordinates of the scalar wavepacket are also not gauge-independent. Hence the necessity to formulate ray-tracing equations on the vectorial wavepacket coordinates.\\
Eqs.~\eqref{eq:zdot_scalar_wavepacket}-\eqref{eq:kzdot_scalar_wavepacket} then yield
\begin{eqnarray}
    \dot {\langle z \rangle}_{\bm{X}} &=& + \partial_{k_z}\tilde{\Omega} + \epsilon F \dot {\langle z \rangle},\label{eq:ray-tracing-1}\\
    \dot {\langle k_z \rangle}_{\bm{X}} &=& - \partial_{z}\tilde{\Omega} + \epsilon F \dot {\langle k_z \rangle},\label{eq:ray-tracing-2}
\end{eqnarray}
where
\begin{eqnarray}
    \tilde{\Omega} &\equiv& \Omega - \frac{i\epsilon}{2} \bm{X}_0^\dagger\{\Omega_0 \mathbf{I}_4,\bm{\chi}_0\} = \Omega_0 + \frac{i\epsilon}{2}
    \bm{\chi}_0^\dagger\{\mathbf{H}_0 - \Omega_0 \mathbf{I}_4,\bm{\chi}_0\} ,\\
    F &=& i\{\bm{\chi}_0^\dagger,\bm{\chi}_0\}.
\end{eqnarray}
$\{a,b\}$ are the Poisson brackets of $a$ and $b$ defined as ${\{a,b\} \equiv \partial_za\,\partial_{k_z}b-\partial_{k_z}a\,\partial_{z}b}$. The full expression of $F$ is given in Appendix~\ref{app:berry-curv} for both acoustic and internal gravity waves.\\

These are the ray-tracing equations of asteroseismology at order $\epsilon^1$ in finite horizontal size. Working with gauge-invariant coordinates in phase space, the system becomes non-canonically Hamiltonian, due to the term $F$ in the right-hand side \citep{littlejohn1991}. Non-canonical Hamiltonian systems are occasionally encountered in fluid mechanics (e.g.\citep{morrison1998}). These systems remain Hamiltonian, and the paths of standing waves remain closed.\\
This demonstrates that the spatial variations of the polarization relations $\bm{\chi}$ affects the ray trajectories in addition to the dispersion relation $\Omega$. This effect is captured in the Berry curvature $F$, which is directly given by the derivatives of $\bm{\chi}$.\\

\section{Berry phase in normal modes}

Equations~\eqref{eq:ray-tracing-1}-\eqref{eq:ray-tracing-2} describe the trajectory of a wavepacket for any given initial condition. The trajectories include those of standing waves in the star, which are the specific solutions that are periodic in time. The trajectory represents the radial oscillatory motion of the perturbation inside the star, bouncing between two turning points. These solutions correspond to closed trajectories $(z(\tau),k_z(\tau))$ in the phase space, traveled in a time $T = 2\pi/\omega$ satisfying the condition
\begin{eqnarray}
    \omega = \Tilde{\Omega}(z,k_z),\label{eq:slice_of_Omega}
\end{eqnarray}
such that the total phase of the wavepacket over one period is an integer multiple of $2\pi$, i.e.
\begin{eqnarray}
    \Delta\phi = 2\pi (n+1) = \frac{1}{\epsilon}\oint_{\Gamma_\omega} (k_z\,\mathrm{d}z + \epsilon i\bm{\chi}_0^\dagger\cdot\mathrm{d} \bm{\chi}_0) + \pi.
\end{eqnarray}
The last term $\pi$ accounts for the two reflections at the turning points of the wave. $\Gamma_\omega$ represents the periodic trajectory in phase space that satisfies $\omega=\Omega(z,k_z)\big|_{(z,k_z)\in\Gamma_\omega}$ oriented clockwise. The integer $n$ is chosen to correspond to the conventional radial order $n$ used in asteroseismology.\\
Applying Stokes' theorem to the second term within the integral, 
one obtains
\begin{eqnarray}
    &&\oint_{\Gamma_\omega} k\,\mathrm{d}z = \epsilon\bigg(2\pi\big(n+\frac{1}{2}\big)+\phi_\mathrm{B}\bigg),\label{eq:Bohr-sommerfeld}
\end{eqnarray}
with
\begin{equation}
    \phi_\mathrm{B} \equiv \iint_{\Sigma_\omega}F\,\mathrm{d}z\mathrm{d}k_z ,
    \label{eq:phib}
\end{equation}
where $\Sigma_\omega$ denotes the area of the phase space enclosed by $\Gamma_\omega$. This procedure, which derives the normal modes from ray dynamics, is known as Bohr-Sommerfeld quantization in quantum physics. It introduces here $\phi_\mathrm{B}$ referred to as the \textit{geometric phase} or the \textit{Berry phase} \citep{berry1984}. This term arises from the vectorial nature of the wave, which perturbs multiple fields simultaneously, with relative phases determined by $\bm{\chi}_0$. In an inhomogeneous medium, the gradual change of this vector along the propagation causes the ray trajectory to bend, in addition to the variation of the dispersion relation $\Omega_0$ \citep{perez2021}. This phenomenon is captured by the terms proportional to the Berry curvature $F$ in the ray-tracing equations~\eqref{eq:ray-tracing-1}-\eqref{eq:ray-tracing-2} or equivalently, as the Berry phase in the frequencies of the normal modes. We provide an explicit expression for $F$ for acoustic and gravity waves in Appendix~\ref{app:berry-curv}.\\

Figure~\ref{fig:soleil} shows that the Berry phase can play a significant role in determining low-order $p$-modes in the Sun. We numerically calculate the frequencies of the normal modes of oscillation for the standard model of the Sun \citep{christensen1996} directly from Eq.~\eqref{eq:schrodinger_equation}, and compare these results with those obtained using Duvall's law, usual scalar theories, and the vectorial Bohr-Sommerfeld quantization that includes $\phi_\mathrm{B}$. Additionally, we compare the results with the available observational helioseismic data from GONG \citep{hill1996}, which includes modes from $n = 4$ to $n = 22$ that have been observed at $\ell=25$.\\

At high radial orders ($n \gtrsim 20$), all methods converge and agree within a relative error of approximately $\sim 1\%$. For low $n$ instead, the prediction accounting for $\phi_\mathrm{B}$ is the only method that achieves a comparable level of accuracy.  Notably, Berry's phase contributes significantly to the total frequency, accounting for $9\%$ of the frequency of the $n=0$ mode, and $7\%$ of the frequency of the $n=1$ mode. For these modes, it is found to be $\phi_\mathrm{B} \simeq -0.71$ and $\phi_\mathrm{B} \simeq -1.17$ respectively. The values of Berry curvature $F$ for acoustic waves in the Sun are shown on Fig~\ref{fig:phase-space}, as well as the $n=0..10$ standing waves in phase space. We stress that a quantitative evaluation of the Berry phase should not assume the Cowling approximation in order to match the high level of modern observational precision.\\

Equation~\eqref{eq:Bohr-sommerfeld} is therefore satisfied by the pulsations $\omega$ of the standing waves for small $\epsilon$, i.e for large degrees $\ell$. In this limit, it gives the frequencies of the pulsations for any order $n$. The ray-tracing equations were derived for wavepackets with small spatial extensions, leaving the possibility that these equations are not guaranteed to hold for waves of low radial order $n$. However, the Bohr-Sommerfeld quantization holds for all $n$, as evidenced by numerical values obtained on Fig.~\ref{fig:soleil}. This effect has been identified in the past. For low $n$, the trajectories in phase space are close to the extremum of $\tilde{\Omega}(z,k_z)$, and as such are following the dynamics of a harmonic oscillator. Since Bohr-Sommerfeld quantization is exact for harmonic oscillators \citep{argyres1965}, the law appears valid in both the high-$n$ and low-$n$ limits, albeit for different reasons. The correction term should then still be Berry’s phase and is still accurate, even though it is not a slow change of the Hamiltonian anymore. We suspect a deeper underlying principle explains why the Bohr-Sommerfeld law applies universally, but this remains unclear at present.\\

\section{Discussion and conclusion}

The Berry phase should not be confused with the so-called \textit{phase function} $\alpha(\omega)$, which modifies Duvall's law by accounting for corrections due to non-zero values of $S$ in the dispersion relation $\Omega_0$, i.e at order $\epsilon^0$. The Berry phase is a first-order term in $\epsilon$ that arises from the polarization relations. Neglecting $\phi_\mathrm{B}$ is equivalent to considering a scalar ray-tracing theory that propagates using the full dispersion relation in accounting for 
$\alpha\left( \omega \right)$.\\

Historically, the Berry phase was studied in system that slowly vary in time. Here, it manifests as the background quantities vary in space crossed by a propagating ray. Our results show that vectorial ray-tracing accounts for this phase, also sometimes called the holonomy \citep{simon1983}, as well as a corrected dispersion relation $\omega = \Tilde{\Omega}$.\\

\begin{figure}
    \centering
    \includegraphics[width=0.6\columnwidth]{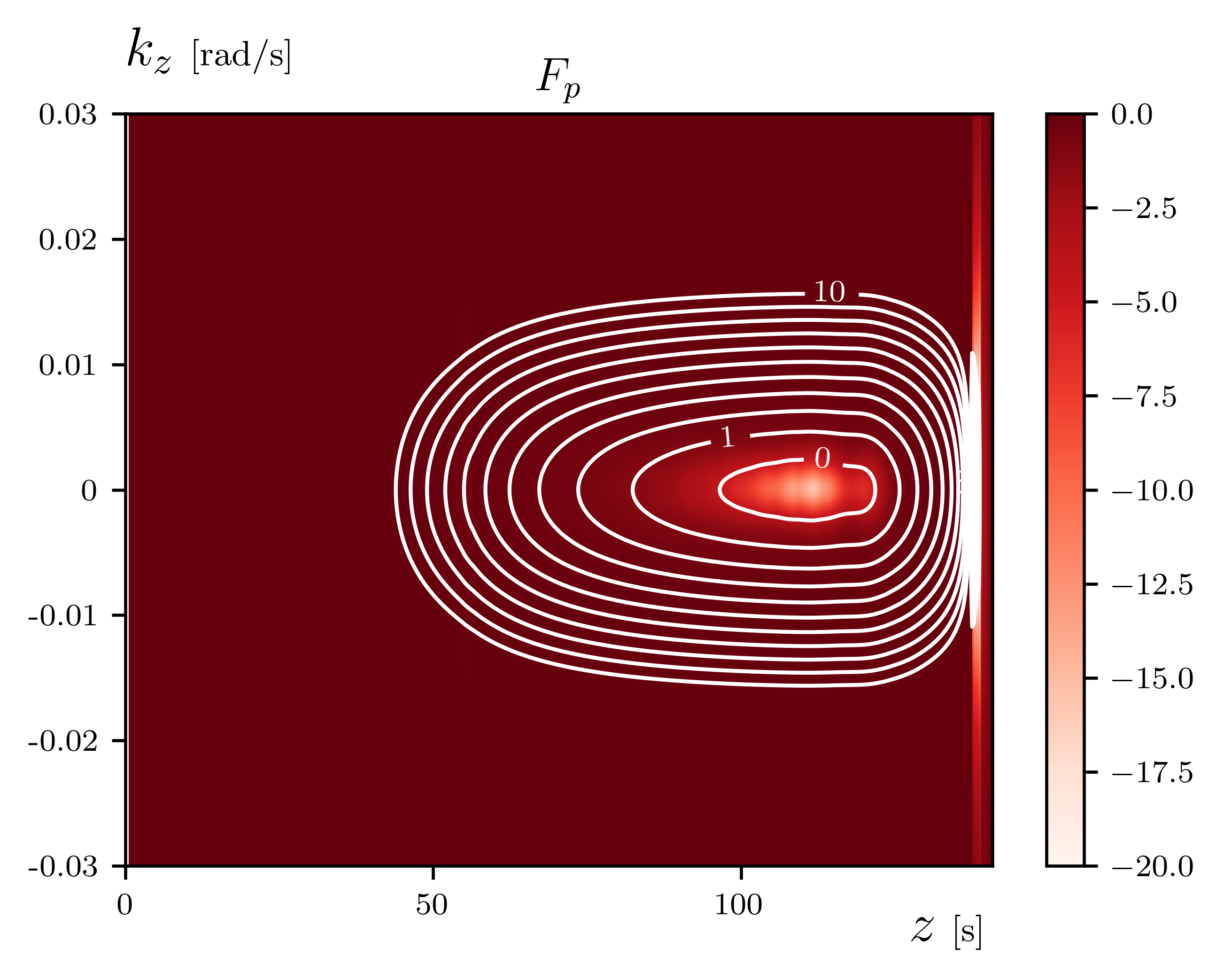}
    \caption{The phase-space trajectories of solar standing acoustic waves enclose a specific amount of Berry curvature $F_{\rm p}$, which gives rise to the Berry phase. A sharp feature appears close to the surface ($z\sim 140$). Radial orders shown vary from $n=0$ to $n = 10$ for $\ell=25$. }
    \label{fig:phase-space}
\end{figure}

The analytical expressions derived in this study enable the prediction of when and 
to what extent effects of geometric phase will be significant in stellar objects. 
\al{We find here that the Berry curvature is significant at the surface as is 
shown on Fig.~\ref{fig:phase-space}, a region where the pressure scale height is 
very short and where several phenomena are poorly modeled such as temperature 
gradients, significant non-adiabaticity and rapid convection \citep{gough1990,ball2014}.}\\
Other stellar situations are known to cause difficulties in normal modes computation, 
among which are found the problem of glitches in red giants \citep{cunha2015}, 
rapidly rotating stars \citep{lignieres2009,mirouh2022} and magnetized stars 
\citep{loi2020}, where complex geometries and anisotropy complicate the study of normal 
modes, but ray-tracing equations remain directly applicable \citep{gough1993}. 
\al{Two extensions should be performed to apply this study on these problems. 
Firstly, to adapt the formalism to low degrees $\ell\lesssim3$ in the spirit of 
\citet{roxburgh2000}. Secondly, to extend the theory for ray-tracing of mixed modes, 
where frequencies of $g$-modes and $p$-modes take similar values, implying for the 
bands to be not well separated as occurs in red giants \citep{mosser2014}.}\\

\al{Ray-tracing equations are successfully employed by local helioseismology, 
which examines point sources and time travels of waves at the solar surface 
(e.g. \citealt{gizon2010}). Our findings suggest that these studies may be 
extended to cases where the $\epsilon^0$ order lacks sufficient accuracy, such 
as for large-scale excitations.\\}

We finally highlight that in this problem, the Berry curvature arises from two Berry-Chern monopoles with topological charges $\mathcal{C}=\pm1$ situated at low $\ell$ values \citep{perrot2019,leclerc2022}. These sources of the Berry curvature impose that $\phi_\mathrm{B}$ tends to $-\vert\mathcal{C}\vert \pi = -\pi$ for trajectories that encompass the entire phase space, which corresponds to large $n$. This amounts to subtracting -1/2 to $n+1/2$ in the quantization law for high $n$.\\ 

\textbf{Acknowledgments:} The authors thank the anonymous referee for a thorough review, and Isabelle Baraffe for discussions on applications. AL is funded by Contrat Doctoral Spécifique Normaliens. GL acknowledges funding from ERC CoG project PODCAST No 864965. The scripts written to support our conclusions are available in the Zenodo dataset \dataset[10.5281/zenodo.14944308]{10.5281/zenodo.14944308}.

\newpage
\bibliographystyle{unsrtnat}

\appendix

\section{Expressions of dispersion relations, polarizations, and Berry curvature}
\label{app:berry-curv}

For acoustic waves, the dispersion relation at order $\epsilon^0$ reads
\begin{equation}
    \omega = \Omega_{0,p}(z,k_z) = \frac{1}{\sqrt{2}}\sqrt{k_z^2 + L_\ell^2 + N^2 + S^2 + \sqrt{(k_z^2 + L_\ell^2 + N^2 + S^2)^2 - 4N^2L_\ell^2}}.
\end{equation}
Their normed polarization relations at order $\epsilon^0$ are
\begin{equation}
    \begin{pmatrix}
        v_h \\ v_r \\ \Theta \\ p
    \end{pmatrix} = \chi_{0,p} = 
    \begin{pmatrix}
        \frac{1}{\Omega_{0,p}}L_\ell\\ 
        \frac{\Omega_{0,p}}{\Omega_{0,p}^2 - N^2}(k_z+iS) \\ 
        \frac{iN}{\Omega_{0,p}^2 - N^2}(k_z+iS) \\ 
        1
    \end{pmatrix}\Bigg/\Bigg|\Bigg|\begin{pmatrix}
        \frac{1}{\Omega_{0,p}}L_\ell\\ 
        \frac{\Omega_{0,p}}{\Omega_{0,p}^2 - N^2}(k_z+iS) \\ 
        \frac{iN}{\Omega_{0,p}^2 - N^2}(k_z+iS) \\ 
        1
    \end{pmatrix}\Bigg|\Bigg|.
\end{equation}
The expression of the Berry curvature for acoustic waves is

\begin{eqnarray}
    F_{\rm p}(z,k_z) &=& \frac{1}{\bigg(\big(k_z^2+(L_\ell-N)^2+S^2\big)\big(k_z^2+(L_\ell+N)^2+S^2\big)\bigg)^{3/2}}\bigg[\nonumber\\
    &&\big(k_z^2+S^2+L_\ell^2+3N^2\big)SL_\ell L_\ell^\prime\nonumber\\
    &-&\big(k_z^2+S^2+3L_\ell^2+N^2\big)SNN^\prime\nonumber\\
    &-&\big(k_z^2+S^2+L_\ell^2+N^2\big)\big(L_\ell^2-N^2\big)S^\prime
    \bigg].\label{eq:express_Berry_curv}
\end{eqnarray}
The three external parameters $N,L_\ell$ and $S$ are functions of $z$ and describe the stratified background. ${}^\prime$ denotes the derivative with respect to $z$.\\

For internal gravity waves, the dispersion relation at order $\epsilon^0$ reads
\begin{equation}
    \omega = \Omega_{0,g}(z,k_z) = \frac{1}{\sqrt{2}}\sqrt{k_z^2 + L_\ell^2 + N^2 + S^2 - \sqrt{(k_z^2 + L_\ell^2 + N^2 + S^2)^2 - 4N^2L_\ell^2}}.
\end{equation}
Their polarization relations at order $\epsilon^0$ are
\begin{equation}
    \begin{pmatrix}
        v_h \\ v_r \\ \Theta \\ p
    \end{pmatrix} = \chi_{0,g} = 
    \begin{pmatrix}
        \frac{1}{\Omega_{0,g}}L_\ell\\ 
        \frac{\Omega_{0,g}}{\Omega_{0,g}^2 - N^2}(k_z+iS) \\ 
        \frac{iN}{\Omega_{0,g}^2 - N^2}(k_z+iS) \\ 
        1
    \end{pmatrix}\Bigg/\Bigg|\Bigg|\begin{pmatrix}
        \frac{1}{\Omega_{0,g}}L_\ell\\ 
        \frac{\Omega_{0,g}}{\Omega_{0,g}^2 - N^2}(k_z+iS) \\ 
        \frac{iN}{\Omega_{0,g}^2 - N^2}(k_z+iS) \\ 
        1
    \end{pmatrix}\Bigg|\Bigg|.
\end{equation}
The Berry curvature of internal gravity waves is given by
\begin{equation}
    F_{\rm g} = -F_{\rm p}.
\end{equation}

Mathematically, there are two additional wavebands: the acoustic and internal gravity waves with negative frequencies, making a total of four. The only differences are that their dispersion relations have the opposite sign of their positive counterparts, while their Berry curvatures remain unchanged.

\end{document}